%% file: maxwell.tex
\documentclass[12pt]{amsart}
\usepackage[dvips,colorlinks=true,linkcolor=blue,urlcolor=red]{hyperref}
\numberwithin{equation}{section}
\newcommand{\smallpagebreak}{{\par\vspace{2 mm}\noindent}}

\newcommand{\R}{\mathbb{R}}

\newcommand{\C}{\mathbb{C}}

\def \al{\alpha}
\def \be{\beta}

\def \de{\delta}
\def \er{\varepsilon}

\def \la{\lambda}

\def \ph{\varphi}

\def \La{\Lambda}
\def \OO{\Omega}

\def \RR{\mathbb{R}}
\def \ZZ{\mathbb{Z}}
\def\dom{\operatorname{Dom}}
\def\div{\operatorname{div}}
\def\mes{\operatorname{mes}}
\def\im{\operatorname{Im}}

\newcommand{\<}{\langle}
\renewcommand{\>}{\rangle}
\theoremstyle{plain}
\newtheorem{Th}{Theorem}[section]
\newtheorem{Le}{Lemma}[section]

\theoremstyle{definition}
\newtheorem{Rem}{Remark}[section]
\title{Absolutely continuous spectrum for the isotropic Maxwell
  operator with coefficients that are periodic in some directions and
  decay in others}
\author{N. Filonov} \address[Nikola{\"\i} Filonov]{Department of
  Mathematical Physics, St Petersburg State University, 1 Ulyanovskaya,
  198504 St Petersburg-Petrodvorets, Russia}
\email{\href{mailto:filonov@mph.phys.spbu.ru}{filonov@mph.phys.spbu.ru}}
\author{F. Klopp} \address[Fr{\'e}d{\'e}ric Klopp]{LAGA, U.M.R. 7539
  C.N.R.S, Institut Galil{\'e}e, Universit{\'e} de Paris-Nord, 99 Avenue J.-B.
  Cl{\'e}ment, F-93430 Villetaneuse, France}
\email{\href{mailto:klopp@math.univ-paris13.fr}{klopp@math.univ-paris13.fr}}
\thanks{N.F.'s research was partially supported by the FNS 2000
  ``Programme Jeunes Chercheurs''. F.K.'s research was partially
  supported by the program RIAC 160 at Universit{\'e} Paris 13 and by the
  FNS 2000 ``Programme Jeunes Chercheurs''.\\
  The authors are grateful to Prof. P.~Kuchment for drawing their
  attention to the question addressed in the present paper}
\begin{document}
\begin{abstract}
  The purpose of this paper is to prove that the spectrum of an
  isotropic Maxwell operator with electric permittivity and magnetic
  permeability that are periodic along certain directions and tending
  to a constant super-exponentially fast in the remaining directions
  is purely absolutely continuous. The basic technical tools is a new
  ``operatorial'' identity relating the Maxwell operator to a
  vector-valued Schr{\"o}dinger operator. The analysis of the spectrum of
  that operator is then handled as in~\cite{Fi-Kl:04,Fi-Kl:04c}.
%% \vskip.5cm\noindent
%% \textsc{R{\'e}sum{\'e}.}

\end{abstract}
\setcounter{section}{-1}
\maketitle
\input{int-max.tex}
%
\input{cal-max.tex}
%
\input{abs-max.tex}

%
\def\cprime{$'$} \def\cydot{\leavevmode\raise.4ex\hbox{.}}

%% \bibliographystyle{plain}
%% \bibliography{biblio}
%
\end{document}

%% file: int-max.tex
\section{The main result}
\label{intro}
In $\R^3$, we study the Maxwell operator
\begin{equation}
  \label{eq:1}
M=i\begin{pmatrix}
  0 & \varepsilon^{-1}\nabla\times\cdot\\-\mu ^{-1}\nabla\times \cdot& 0
\end{pmatrix}
\end{equation}
acting on the space $\mathcal{H}(\varepsilon)\oplus\mathcal{H}(\mu )$.
Here, $\nabla$ denotes the gradient of a function, $\div$ the
divergence of a vector field, $\times$ the standard cross-product in
$\R^3$, and we defined
\begin{equation*}
  \mathcal{H}(\varepsilon):=\{u\in L^2(\R^3,\varepsilon(x)
  dx)\otimes\C^3;\ \div(\varepsilon x)=0\}.
\end{equation*}
$\mathcal{H}(\varepsilon)$ is endowed with its natural scalar product
\begin{equation*}
  \langle f,g\rangle_\varepsilon=\int_{\R^3}\langle
  f(x),g(x)\rangle_\C\,\varepsilon(x)dx
\end{equation*}
where $\langle\cdot,\cdot\rangle_\C$ denotes the usual scalar product
in $\C^3$.\\
Pick $d\in\{1,2\}$. Let $(x,y)$ denote the points of the space
$\mathbb{R}^3$. Define $\OO=\RR^{3-d}\times (0, 2\pi)^d$.
\vskip.2cm\noindent We assume that the scalar functions $\varepsilon$
and $\mu $ satisfy
\begin{description}
\item[(H1)] $\forall l\in\ZZ^d$, $\forall(x,y)\in\R^3$,
  \begin{equation*}
    \varepsilon(x,y+2\pi l)=\varepsilon(x,y),\quad \mu (x,y+2\pi l)=\mu (x, y);
  \end{equation*}
\item[(H2)] the functions $\varepsilon$ and $\mu $ are twice continuously
  differentiable in $\Omega$;
\item[(H3)] there exist $\varepsilon_0>0$ and $\mu _0>0$ such that, for
  any $a>0$, one has
  \begin{equation*}
   \sup_{0\leq|\alpha|\leq2}\sup_{(x,y)\in\Omega}e^{a|x|}
   (|\partial^{\alpha}(\varepsilon-\varepsilon_0)(x,y)|+
   |\partial^{\alpha}(\mu -\mu _0)(x,y)|)<+\infty ;
  \end{equation*}
\item[(H4)] there exists $c_0>0$ such that $\forall(x,y)\in\R^3$,
  $\varepsilon(x,y)\geq c_0$ and $\mu (x,y)\geq c_0$.
\end{description}
Then, our main result is
\begin{Th}
  \label{thr:1}
  Under assumptions (H1)--(H4), the spectrum of $M$ is purely
  absolutely continuous.
\end{Th}
\noindent In~\cite{MR2001h:82103}, A.~Morame proved that the spectrum
of the Maxwell operator~\eqref{eq:1} is absolutely continuous when the
electric permittivity $\varepsilon$ and the magnetic permeability $\mu $
are periodic with respect to a non-degenerate lattice in $\R^3$.
In~\cite{MR2003e:35224}, T.~Suslina proved the absolute continuity of
the spectrum of the Maxwell operator~\eqref{eq:1} in a strip when the
electric permittivity $\varepsilon$ and the magnetic permeability $\mu $
are periodic along the strip (with perfect conductivity conditions
imposed on the boundary of the strip). \\
In both papers, the authors first apply a standard idea in the
spectral theory of the Maxwell operator to circumvent one of the first
technical difficulties one encounters when dealing with the Maxwell
system: the fact that the domain of the Maxwell operator,
$\mathcal{H}(\varepsilon)\oplus\mathcal{H}(\mu )$, consists of only the
divergence free vectors (up to multiplication by $\varepsilon$ or
$\mu $). To resolve that difficulty, the standard
idea~\cite{BS} is to extend the Maxwell
operator to an operator acting on $L^2(\R^3)\otimes\C^8$. We introduce
such an extension that slightly differs from the one considered
in~\cite{BS,MR2002k:78002,MR2003e:35224,MR2001h:82103} as we
require some additional properties.
\smallpagebreak Consider the matrix of first order linear
differential expressions
\begin{equation}
    \label{eq:4}
  {\mathcal M}=i\begin{pmatrix}
    0 & \varepsilon^{-1}\nabla\times \cdot & 0 & \nabla(\varepsilon^{-1}\,\cdot )
    \\-\mu ^{-1}\nabla\times \cdot & 0 & \nabla(\mu ^{-1}\,\cdot ) &0 \\
    0 & (\varepsilon\mu )^{-1}\div(\mu \,\cdot ) & 0 & 0
    \\ (\varepsilon\mu )^{-1}\div(\varepsilon\,\cdot ) & 0 &0 &0
  \end{pmatrix}.
\end{equation}
It naturally defines an elliptic self-adjoint operator on
\begin{equation*}
  \mathcal{H}_{\text{tot}}:=L^2(\R^3,\varepsilon(x)dx;\C^3)
  \oplus L^2(\R^3,\mu (x)dx;\C^3)
  \oplus L^2(\R^3,\varepsilon(x)dx)\oplus L^2(\R^3,\mu (x)dx)
\end{equation*}
with domain
\begin{equation*}
  H^1(\R^3;\C^3)\oplus H^1(\R^3;\C^3)
  \oplus H^1(\R^3)\oplus H^1(\R^3).
\end{equation*}
Let $\Pi$ be the orthogonal projector on
$\mathcal{H}(\varepsilon)\oplus\mathcal{H}(\mu )\oplus\{0\}\oplus\{0\}$
in $\mathcal{H}_{\text{tot}}$. One checks that
\begin{equation}
  \label{eq:37}
  [\Pi,\mathcal{M}]=0.
\end{equation}
This is a consequence of the well known facts that gradient fields are
orthogonal (for the standard scalar product) to divergence free
fields, and that curl fields are divergence free.\\
Moreover, one computes
\begin{equation}
  \label{part}
  \Pi\mathcal{M}\Pi=\Pi\begin{pmatrix} M & 0\\0 &0\end{pmatrix}\Pi.
\end{equation}
This and equation~\eqref{eq:37} imply that Theorem~\ref{thr:1} is an
immediate consequence of
\begin{Th}
  \label{thr:2}
  Under assumptions(H1)--(H4), the spectrum of $\mathcal{M}$ is purely
  absolutely continuous.
\end{Th}
\noindent In the cases dealt with
in~\cite{MR2001h:82103,MR2003e:35224}, to prove the absolute
continuity of the spectrum of $\mathcal{M}$ (or rather said their
analogue of $\mathcal{M}$), the authors perform the
Bloch-Floquet-Gelfand reduction that brings them back to studying an
operator with compact resolvent. Because of this, they only need to
show that $\mathcal{M}$ has no eigenvalue. To prove this, they show
that the fact that $\mathcal{M}$ has an eigenvalue implies that some
Schr{\"o}dinger operator with a potential having the same symmetry
properties as $\varepsilon$ and $\mu $ has an eigenvalue.  The well known
argument showing that this is impossible relies on the fact that the
reduced operator has compact resolvent.
\smallpagebreak In our case, by assumption (H1), the
Bloch-Floquet-Gelfand reduction can only be done in the $y$-variable;
hence, the resolvent of the reduced operator is not compact. So, the
standard argument does not apply. To analyze the reduced
$\mathcal{M}$, we first show an ``operatorial'' identity that brings
us back to analyzing a Schr{\"o}dinger operator; then, to analyze this
Schr{\"o}dinger operator, we apply the method developed
in~\cite{Fi-Kl:04}.
\smallpagebreak Consider the following differential
matrices acting on twice differentiable functions valued in
$\C^3\oplus\C^3\oplus\C\oplus\C$
\begin{gather}
  \label{eq:8}
  \Delta_8:=
  \begin{pmatrix}\Delta_3 & 0 &0 &0\\0 & \Delta_3 &0 &0 \\ 0& 0 &
    \Delta &0\\ 0 & 0 & 0 & \Delta\end{pmatrix}\quad\text{where}\quad
    \Delta_3:=\begin{pmatrix}\Delta & 0 &0\\0 & \Delta &0 \\ 0& 0 &
    \Delta\end{pmatrix},\\
  \label{eq:5}
  {\mathcal A}=i\begin{pmatrix} 0 & -\mu z\times \cdot & 0 & -\mu z\,\cdot
    \\\varepsilon z\times \cdot & 0 & -\varepsilon z\,\cdot &0 \\
    0 & 0 & 0 & 0 \\ 0 & 0 &0 &0
  \end{pmatrix},\\
  \label{eq:6}
  {\mathcal J}=\begin{pmatrix} \varepsilon^{-1/2}\,\cdot &0 & 0 &0
    \\ 0& \mu ^{-1/2}\,\cdot &0 &0 \\
    0 &0 & \mu ^{1/2}\,\cdot & 0 \\ 0 & 0 &0 & \varepsilon^{1/2}\,\cdot
  \end{pmatrix}
\end{gather}
where $\Delta$ is the standard Laplace operator in $\R^3$ and
\begin{equation}
  \label{eq:3}
  z=\nabla((\varepsilon\mu )^{-1}).
\end{equation}
We prove
\begin{Th}
  \label{thr:3}
  One computes
  \begin{equation}
    \label{eq:7}
    \varepsilon\mu \mathcal{J}^{-1}(\mathcal{M}+
    \mathcal{A})\mathcal{M}\mathcal{J}=
    -\Delta_8+\mathcal{V}+\mathcal{F}
  \end{equation}
  where
  \begin{itemize}
  \item $\Delta_8$ is the diagonal Laplace operator defined
    in~\eqref{eq:8},
  \item $\mathcal{V}$ is the zeroth-order matrix and $\mathcal{F}$ the
    first-order matrix defined by
    \begin{equation}
      \label{eq:9}
      \mathcal{V}=\begin{pmatrix} V(\varepsilon)\cdot &0 & 0 &0
        \\ 0& V(\mu )\cdot &0 &0 \\ 0 &0 & v(\mu )\cdot & 0 \\ 0 & 0 &0 &
        v(\varepsilon)\cdot \end{pmatrix},\quad
      \mathcal{F}=\begin{pmatrix}
        0  &0 & -F(\varepsilon,\mu ,\cdot ) &0 \\ 0& 0 &0 &
        F(\mu ,\varepsilon,\cdot ) \\ 0 &0 & 0 & 0 \\ 0 & 0 &0 &0
      \end{pmatrix},
    \end{equation}
    and, for $\{f,g\}=\{\mu ,\varepsilon\}$, we have defined
    \begin{gather}
      \label{eq:10}
      V(f)=v(f)\text{Id}-2\text{Jac}(s(f)),\quad v(f)=s^2(f)+\div s(f)
      \quad\text{and}\quad s(f)=f^{-1/2}\nabla(f^{1/2}) ,\\
      \label{eq:11}
      F(f,g,\cdot )=f^{-1/2}\nabla(\varepsilon \mu )\times \nabla(g^{-1/2}\,\cdot ),
    \end{gather}
    and Jac$(g)$ denotes the Jacobian of a differentiable function $g:\
    \R^3\to\R^3$.

  \end{itemize}
\end{Th}
\begin{Rem}
  If the functions $\varepsilon$, $\mu$ are such that the product
  $\varepsilon\mu $ is constant then $\mathcal{A} = 0$ and $\mathcal{F} =
  0$.
  This idea was used in \cite{Fil}.
\end{Rem}
\begin{Rem}
  Though computations analogous to those leading to
  Theorem~\ref{thr:3} have been done
  in~\cite{MR2001h:82103,MR2003e:35224}, to our knowledge, the
  ``operatorial'' identity~\eqref{eq:7} is new. We hope it will also
  prove useful beyond the present study~\cite{Fi-Kl:04b}.
\end{Rem}
\begin{Rem}
  As a consequence of~\eqref{eq:7}, for $\lambda\in\C$, we obviously
  obtain
  \begin{equation}
    \label{eq:13}
    \varepsilon\mu \mathcal{J}^{-1} (\mathcal{M}+\mathcal{A}+\lambda)
    (\mathcal{M} - \lambda) \mathcal{J}
    = -\Delta_8+\mathcal{V} - \varepsilon \mu \mathcal{J}^{-1}
    (\lambda \mathcal{A} + \lambda^2)\mathcal{J} + \mathcal{F}.
  \end{equation}
  These equalities being written between differential matrices can be
  complemented with boundary conditions to yield equalities between
  operators. Among the boundary conditions we will need are the
  quasi-periodic Floquet boundary conditions described in
  section~\ref{sec:proof-theor-refthr:1}.
\end{Rem}
\begin{Rem}
One can consider another extension of the initial operator (\ref{eq:1}),
\begin{equation*}
  {\mathcal M}=i\begin{pmatrix}
    0 & \varepsilon^{-1}\nabla\times \cdot &
  0 & \nabla(\al_2\be_2 \,\cdot )
    \\-\mu ^{-1}\nabla\times \cdot & 0 &
  \nabla(\al_1\be_1\,\cdot ) &0 \\
    0 & \beta_1\div(\mu \,\cdot ) & 0 & 0
    \\ \beta_2\div(\varepsilon\,\cdot ) & 0 &0 &0
  \end{pmatrix}
\end{equation*}
with positive functions $\al_1, \al_2, \be_1, \be_2$.  This operator is
self-adjoint in the space
\begin{equation*}
  L^2(\R^3,\varepsilon(x)dx;\C^3) \oplus L^2(\R^3,\mu (x)dx;\C^3)
  \oplus L^2(\R^3,\al_1(x)dx)\oplus L^2(\R^3,\al_2 (x)dx)
\end{equation*}
and (\ref{part}) holds.  If $\al_1 \be_1^2 = \varepsilon^{-1}
\mu^{-2}$ and $\al_2 \be_2^2 = \varepsilon^{-2} \mu^{-1}$ and we take
\begin{equation*}
  {\mathcal A}=i\begin{pmatrix} 0 & -\mu z\times \cdot &
 0 & -\be_2^{-1} \varepsilon^{-1} z\,\cdot
    \\\varepsilon z\times \cdot &
 0 & -\beta_1^{-1} \mu^{-1} z\,\cdot &0 \\
    0 & 0 & 0 & 0 \\ 0 & 0 &0 &0
  \end{pmatrix},\\
\end{equation*}
and ${\mathcal J}= diag(\varepsilon^{-1/2}, \mu ^{-1/2}, \al_1^{-1}
\be_1^{-1} \mu ^{-1/2}, \al_2^{-1} \be_2^{-1} \varepsilon^{-1/2})$ then
formulae (\ref{eq:7}), (\ref{eq:13}) still hold (our choice in this
paper is $\al_1 = \varepsilon$, $\al_2 = \mu$, $\beta_1 = \beta_2 =
\varepsilon^{-1}\mu^{-1}$).
\end{Rem}

%%% Local Variables:
%%% mode: latex
%%% TeX-master: "maxwell"
%%% End: 

%% file: cal-max.tex
\section{A useful formula: the proof of Theorem~\ref{thr:3}}
\label{sec:useful-formula}
The computations leading to Theorem~\ref{thr:3} are quite similar to
those done in~\cite{MR2003e:35224}.\\
We first compute
\begin{equation*}
     \mathcal{M}\mathcal{J}=
     i\begin{pmatrix} 0 &\varepsilon^{-1}\nabla\times (\mu^{-1/2}\,\cdot )
      & 0 & \nabla(\varepsilon^{-1/2}\,\cdot )
      \\-\mu^{-1}\nabla\times (\varepsilon^{-1/2}\,\cdot )& 0 &
      \nabla(\mu^{-1/2}\,\cdot ) &0 \\
      0 & (\varepsilon\mu)^{-1}\div(\mu^{1/2}\cdot ) & 0 & 0 \\
      (\varepsilon\mu)^{-1}\div(\varepsilon^{1/2}\,\cdot ) & 0 &0 &0
    \end{pmatrix}.
\end{equation*}
Hence, as $\div(\nabla\times \cdot )=0$ and $\nabla\times \nabla\cdot =0$, we obtain
\begin{equation}
  \label{eq:22}
  \varepsilon\mu \mathcal{J}^{-1}\mathcal{M}^2\mathcal{J}=-
  \begin{pmatrix}
    a(\varepsilon,\mu) & 0 & 0 & 0\\
    0 & a(\mu ,\varepsilon) & 0 & 0\\
    0 & 0 & b(\mu ) &0\\ 0& 0 & 0 & b(\varepsilon)
  \end{pmatrix}
\end{equation}
where, for $\{f,g\}=\{\varepsilon,\mu \}$, we have defined
\begin{gather}
  \label{eq:12}
  a(f,g)=-f^{1/2}g\nabla\times  (g^{-1}\nabla\times (f^{-1/2}\,\cdot ))+
  (fg)f^{1/2}\nabla(f^{-1}(fg)^{-1}\div(f^{1/2}\cdot )),\\
  \label{eq:14}
  b(f)=f^{-1/2}\div(f\nabla(f^{-1/2}\,\cdot )).
\end{gather}
On the other hand,
\begin{equation}
  \label{eq:21}
    \varepsilon\mu \mathcal{J}^{-1}\mathcal{A}\mathcal{M}\mathcal{J}=-
    \begin{pmatrix} c(\varepsilon) & 0 &-d(\mu ) &0 \\ 0& c(\mu )& 0 &
      d(\varepsilon)\\ 0 & 0 & 0 & 0 \\ 0 & 0 &0 &0
    \end{pmatrix}.
\end{equation}
where, for $f\in\{\varepsilon,\mu \}$, we have defined
\begin{gather}
  \label{eq:16}
  c(f)=\varepsilon\mu (f^{1/2}z\times \nabla\times (f^{-1/2}\,\cdot )
  -zf^{-1/2}\div(f^{1/2}\cdot )),\\
  \label{eq:17}
  d(f)=(\varepsilon \mu )^{3/2}f^{1/2}z\times \nabla(f^{-1/2}\,\cdot ).
\end{gather}
For $\{f,g\}=\{\varepsilon,\mu \}$, using~\eqref{eq:17} and
\begin{equation}
  \label{eq:27}
  f\nabla(f^{-1})=-f^{-1}\nabla f,
\end{equation}
we compute
\begin{equation}
  \label{eq:19}
  d(f)=-(\varepsilon \mu )^{-1/2}f^{1/2}\nabla(\varepsilon
  \mu )\times \nabla(f^{-1/2}\,\cdot )=-g^{-1/2}\nabla(\varepsilon\mu )\times \nabla(f^{-1/2}\,\cdot )
  =-F(g,f,\cdot)
\end{equation}
which gives formula~\eqref{eq:11} for the coefficient of the
matrix $\mathcal{F}$ in Theorem~\ref{thr:3}.\\
Recall that, for $u:\ \R^3\to\R$ and $v:\ \R^3\to\R^3$ both once
differentiable, one has
\begin{equation}
  \label{eq:24}
  \nabla\times (u v)=u\,(\nabla\times v)+(\nabla u)\times v.
\end{equation}
Using this,~\eqref{eq:12},~\eqref{eq:16} and~\eqref{eq:3}, we compute
\begin{gather*}
  \begin{split}
    -\varepsilon^{1/2}\mu \nabla\times  ((\varepsilon
    \mu )^{-1}\varepsilon\nabla\times (\varepsilon^{-1/2}\,\cdot ))&+
    \varepsilon^{3/2} \mu \nabla ((\varepsilon \mu )^{-1})
    \times \nabla\times (\varepsilon^{-1/2}\,\cdot)\\&\hskip3cm =
    -\varepsilon^{-1/2}\nabla\times (\varepsilon\nabla\times
    (\varepsilon^{-1/2}\,\cdot )),
  \end{split}
  \\\intertext{and}
   \varepsilon^{3/2} \mu \nabla ((\varepsilon \mu )^{-1}
   \varepsilon^{-1} \div (\varepsilon^{1/2}\,\cdot )) -
   \varepsilon^{1/2}\mu \nabla ((\varepsilon \mu )^{-1})\div(\varepsilon^{1/2}\,\cdot )
   =\varepsilon^{1/2}\nabla(\varepsilon^{-1}\div(\varepsilon^{1/2}\,\cdot )),
\end{gather*}
so
\begin{equation}
  \label{ca}
    c(\varepsilon)+a(\varepsilon,\mu )=
   -\varepsilon^{-1/2}(\nabla\times (\varepsilon\nabla\times
   (\varepsilon^{-1/2}\,\cdot )))+
    \varepsilon^{1/2}\nabla(\varepsilon^{-1}\div(\varepsilon^{1/2}\,\cdot )).
\end{equation}
To complete the proof of Theorem~\ref{thr:3},
taking~\eqref{eq:22},~\eqref{eq:14},~\eqref{eq:21} and~\eqref{ca} into
account, we are only left with proving the following
\begin{Le}
  \label{le:1}
  One has
  \begin{gather}
    \label{eq:15}
    -\varepsilon^{-1/2}(\nabla\times (\varepsilon\nabla\times
    (\varepsilon^{-1/2}\,\cdot )))+
    \varepsilon^{1/2}\nabla(\varepsilon^{-1}\div(\varepsilon^{1/2}\,\cdot ))=
    \Delta_3-V(\varepsilon)\\
    \intertext{and}
    \label{eq:20}
    \varepsilon^{-1/2}\div(\varepsilon\nabla(\varepsilon^{-1/2}\,\cdot ))
    =\Delta-v(\varepsilon)
  \end{gather}
  where $V$ and $v$ are defined in Theorem~\ref{thr:3}.
\end{Le}
\noindent{\it Proof.} We start with the proof
of~\eqref{eq:20}. Using~\eqref{eq:27} and \eqref{eq:10}, we compute
\begin{equation*}
  \begin{split}
    \varepsilon^{-1/2}\div(\varepsilon^{1/2}(\nabla\,\cdot) -
    s(\varepsilon) \varepsilon^{1/2}\,\cdot) &=\Delta + \<s(\er),
    \nabla \cdot \> - \div s(\er) -
    \<s(\er), \varepsilon^{-1/2}\nabla (\varepsilon^{1/2}\,\cdot)\>\\
    &= \Delta - (\div s(\er) + s(\er)^2),
  \end{split}
\end{equation*}
where $\langle\cdot ,\cdot \rangle$ denotes the standard scalar product in
$\R^3$.\\
Let us now prove~\eqref{eq:15}. Using~\eqref{eq:24} we compute
\begin{equation}
  \begin{split}
  \label{250}
  \varepsilon^{-1/2}(\nabla\times &(\varepsilon\nabla\times
  (\varepsilon^{-1/2}\,\cdot )))\\&=
  \varepsilon^{-1/2}\nabla\times (\varepsilon^{1/2}\nabla\times \cdot -
  \varepsilon^{1/2} s(\varepsilon)\times \cdot ) \\&=(\nabla\times \cdot )^2+
  s(\varepsilon)\times (\nabla\times \cdot ) - s(\varepsilon)\times (s(\varepsilon)\times \cdot) -
  \nabla\times (s(\varepsilon)\times \cdot).
  \end{split}
\end{equation}
The classical formula gives
\begin{equation}
  \label{251}
    \nabla\times (s(\varepsilon)\times \cdot) =
    s(\varepsilon) \div\,\cdot - \cdot \div s(\varepsilon)
    - \< s(\varepsilon), \nabla\> \cdot + \< \cdot , \nabla\> s(\varepsilon).
\end{equation}
For the second term in~\eqref{eq:15} we have
\begin{equation}
  \begin{split}
  \label{260}
  \varepsilon^{1/2}\nabla(\varepsilon^{-1}\div(\varepsilon^{1/2}\,\cdot ))&=
  \varepsilon^{1/2}\nabla(\varepsilon^{-1/2}\div(\cdot )+
  \varepsilon^{-1/2}\< s(\varepsilon),\cdot \rangle)\\
  &=\nabla(\div\cdot ) - s(\varepsilon)\div(\cdot )
  - s(\varepsilon)\< s(\varepsilon), \cdot \>
  + \nabla \< s(\varepsilon), \cdot \>.
  \end{split}
\end{equation}
Summarizing~\eqref{250},~\eqref{251} and~\eqref{260} we obtain
\begin{equation*}
  \begin{split}
    -\varepsilon^{-1/2}(\nabla\times
    (\varepsilon\nabla\times (\varepsilon^{-1/2}\,\cdot )))&+
    \varepsilon^{1/2}\nabla(\varepsilon^{-1}\div(\varepsilon^{1/2}\,\cdot ))\\
    &= \Delta_3 - (|s(\varepsilon)|^2 + \div s(\varepsilon))\,\cdot\\
    &\hskip2cm- s(\varepsilon)\times (\nabla\times \cdot ) - \< s(\varepsilon),
    \nabla\> \cdot + \<\,\cdot , \nabla\> s(\varepsilon) + \nabla \<
    s(\varepsilon), \cdot \>,
  \end{split}
\end{equation*}
where the well-known formulas
\begin{equation*}
   \nabla\div - (\nabla\times)^2 = \Delta_3
\end{equation*}
and
\begin{equation*}
   s(\varepsilon)\< s(\varepsilon), \cdot \>
   - s(\varepsilon)\times (s(\varepsilon)\times \cdot)
   = |s(\varepsilon)|^2 \cdot
\end{equation*}
are used.  Now the simple calculations
\begin{gather*}
   s(\varepsilon)\times (\nabla\times \cdot \,) + \< s(\varepsilon), \nabla\>\,\cdot
   = \text{Jac}(\,\cdot \,) s(\varepsilon), \\
   \<\, \cdot \,, \nabla\> s(\varepsilon) = \text{Jac}(s(\varepsilon))^t\,
   \cdot, \quad \nabla \< s(\varepsilon), \cdot \> = \text{Jac}(\,\cdot \,)
   s(\varepsilon) + \text{Jac}(s(\varepsilon))\, \cdot
\end{gather*}
complete the proof of Lemma~\ref{le:1}.\qed
%%% Local Variables:
%%% mode: latex
%%% TeX-master: "maxwell"
%%% End: 

%% file: abs-max.tex
\section{Proof of Theorem~\ref{thr:2}}
\label{sec:proof-theor-refthr:1}
In our previous work~\cite{Fi-Kl:04,Fi-Kl:04c}, we proved the absolute
continuity of the spectrum of the Schr{\"o}dinger operator where the
properties of the potential were similar to those imposed on
permittivity $\varepsilon$ and the permeability $\mu$ in
Theorem~\ref{thr:1} and~\ref{thr:2}. The scheme of the proof of
Theorem~\ref{thr:2} is globally the same as that of Theorem~1.1
in~\cite{Fi-Kl:04,Fi-Kl:04c}; so, we will omit some details.
\smallpagebreak First, basing on the relation (\ref{eq:13}), we
construct a convenient representation of the resolvent $({\mathcal M}
-\la)^{-1}$ (see Lemma~\ref{Lem2.3} below).
\smallpagebreak First of all we need to define some notations. Let
$\<x\>= \sqrt{x^2+1}$. For $a\in\R$, introduce the spaces
\begin{equation*}
  L_{p,a}=\{f: e^{a\<x\>}f\in L_p(\OO)\},\qquad
  H^l_a=\{f:e^{a\<x\>}f\in H^l(\OO)\},
\end{equation*}
where $1\le p\le \infty$ and $H^l(\OO)$ is the standard Sobolev space.
Introduce the function spaces in $\Omega$ with quasi-periodic boundary
conditions
\begin{equation*}
  H^l_a(k):=\left\{f\in H^l_a:(D^\alpha f)\mid_{y_j=2\pi}=e^{2\pi
  ik_j}(D^\alpha f)\mid_{y_j=0},\quad|\alpha|\le l-1\right\}\text{ and
  } H^l(k):=H^l_0(k).
\end{equation*}
Finally, for $X$ and $Y$ Banach spaces, $B(X,Y)$ is the space of all
bounded operators from $X$ to $Y$, and $B(X)=B(X,X)$, both endowed
with their natural norm topology.
\smallpagebreak Due to the Bloch-Floquet-Gelfand transformation, the
Maxwell operator ${\mathcal M}$ is unitary equivalent to the direct
integral $\int_{[0, 1)^d}^\oplus {\mathcal M} (k) dk$, where
${\mathcal M}$ is the operator given by the differential expression
(\ref{eq:4}) on the domain $\dom {\mathcal M} (k) = H^1(k)$.  The
Laplace operator on the
domain $H^2(k)$ will be denoted by $\Delta(k)$.\\
In~\cite{Fi-Kl:04,Fi-Kl:04c}, we essentially proved the following result
\begin{Le}
  \label{Lem2.1}
  Assume that the pair $(k_0, \la_0)\in \RR^{d+1}$ satisfies
  \begin{equation}
    \label{eq:2.1}
    (k_0 + n)^2 \not= \varepsilon_0 \mu_0 \la_0, \qquad \forall n \in \ZZ^d.
  \end{equation}
  Then, there exist numbers $\delta>0$, $a>0$, an open set
  $\Xi_0 \subset \C^{d+1}$ such that
  \begin{equation*}
    \left( B_\de (k_0) \cup \{ k(\tau) \}_{\tau
      \in \RR} \right) \times B_\de (\la_0) \subset \Xi_0,
  \end{equation*}
  where $B_\de (k_0)$ is a ball in real space
  \begin{equation*}
    B_\de (k_0) = \{ k \in \RR^d : |k - k_0| < \de \},
  \end{equation*}
  and $k (\tau) = (\tilde k_1 + i \tau, \tilde k')$ with fixed $\tilde
  k \in B_\de (k_0)$, and there exists an analytic
  $B(L_{2,a},H^2_{-a})$-valued function $R_0$, defined in $\Xi_0$,
  having the properties
  \begin{itemize}
  \item for $(k,\lambda)\in\Xi_0$, $k\in\R^d$, $\im\lambda>0$, $U\in
    L_{2,a}$, one has
    \begin{equation*}
      R_0(k,\lambda)U=(-\Delta(k)-\varepsilon_0\mu_0\lambda)^{-1}U;
    \end{equation*}
  \item
    \begin{equation}
      \label{eq:2.2}
      \|R_0(k(\tau),\lambda)\|_{B(H^2_a,\,H^2_{-a})}\leq C|\tau|^{-1};
    \end{equation}
  \item $R_0(k,\lambda)L_{2, a}\subset H^2_{-a}(k)$.
\end{itemize}
\end{Le}
\noindent This lemma is proved in~\cite{Fi-Kl:04} (see Theorem 3.1)
except for the fact that estimate~\eqref{eq:2.2} is replaced with
\begin{equation}
  \label{eq:2}
  \|R_0(k(\tau),\lambda)\|_{B(L_{2, a},L_{2, -a})}\leq C|\tau|^{-1}.
\end{equation}
The proof of estimate~(\ref{eq:2.2}) is exactly the same as that
of~\eqref{eq:2}.
\smallpagebreak Clearly, in Lemma~\ref{Lem2.1}, we can replace
$\Delta(k)$ with $\Delta_8(k)$ (defined in~\eqref{eq:8}) at
the expense of changing the constants; the resolvent of
$\Delta_8(k)$ (and its analytic extension) will henceforth
be denoted by $R_{\mathcal{M}}^0(k,\la)$. So
\begin{equation*}
  R_{\mathcal{M}}^0(k,\la)=R_0(k,\la)\,\text{Id}_{\C^8}.
\end{equation*}
\smallpagebreak To deal with the potential, we prove
\begin{Le}
  \label{Lem2.2}
  Let $\varepsilon$, $\mu$ satisfy hypothesis (H1)-(H4), ${\mathcal
    A}$ be defined by (\ref{eq:5}), and $(k_0, \lambda_0)$ satisfy
  (\ref{eq:2.1}). Then, there exist $\delta>0$, $a>0$, an open set
  $\Xi \subset \C^{d+1}$ with $B_\de (k_0) \times B_\de (\lambda_0) \subset
  \Xi$, a function $h:\ \Xi\to\C$ analytic in $\Xi$ with the property
  \begin{equation}
    \label{eq:2.3}
    \forall\lambda\in B_\delta(\lambda_0),\quad\exists k\in B_\delta(k_0)\quad
    \text{such that}\quad h(k,\lambda)\not= 0,
  \end{equation}
  and there exists an analytic $B(L_{2,a},H^2_{-a})$-valued function
  $Z$, defined in
  \begin{equation*}
    \Xi_1 := \{ (k, \lambda) \in \Xi : h (k, \lambda)\not= 0\},
  \end{equation*}
  such that, for $(k,\lambda)\in\Xi_1$, $k\in\R^d$, $\im\lambda^2>0$,
  $U\in H^2_a(k)$, one has
  \begin{equation}
    \label{eq:2.4}
    Z(k, \lambda) \left(-\Delta_8 + \mathcal{V} -
      \varepsilon\mu\mathcal{J}^{-1}
      (\lambda \mathcal{A} + \lambda^2)\mathcal{J}\right) U = U
  \end{equation}
  and
  \begin{equation*}
    Z(k, \lambda) L_{2, a} \subset H^2_{-a} (k).
  \end{equation*}
\end{Le}
\begin{proof} Note that
  \begin{equation*}
    \mathcal{V} - \varepsilon\mu\mathcal{J}^{-1}
    (\lambda \mathcal{A} + \lambda^2)\mathcal{J} =
    -\er_0 \mu_0 \la^2 + \mathcal{W} (\la),
  \end{equation*}
  where, by assumptions (H2)-(H3), $\lambda\mapsto{\mathcal
    W}(\lambda)$ is an entire function valued in $L_{\infty,b}$ for
  any $b\in\R$. Set
  \begin{equation*}
    Z(k,\lambda)=\left(I+R_{\mathcal{M}}^0(k,\la^2)\mathcal{W}(\la)
    \right)^{-1}R_{\mathcal{M}}^0(k,\la^2).
  \end{equation*}
  The operator of multiplication by ${\mathcal W}$ is bounded as an
  operator from $H^2_{-a}$ to $H^2_a$, and is compact as an operator
  from $H^2_{-a}$ to $L_{2, a}$. It remains to use the estimation
  (\ref{eq:2.2}) and the analytic Fredholm alternative in the Hilbert
  space $H^2_{-a}$ (see e.g.~\cite{Ka:80,MR58:12429c}) to complete the
  proof of Lemma~\ref{Lem2.2}.
\end{proof}
\noindent In the following lemma, we construct an analytic extension
of the resolvent of Maxwell operator to the non-physical sheet. Set
\begin{equation*}
  Q(\la) = \varepsilon\mu\mathcal{J}^{-1}
(\mathcal{M} + \mathcal{A} + \lambda).
\end{equation*}
Then, for any $b\in\R$, $Q$ is an entire function with values in
$B(H^1_b,L_{2, b})$.
The next result we need is
\begin{Le}
  \label{Lem2.3}
  Under the assumptions of Lemma~\ref{Lem2.2}, on the set $\Xi_1$, we
  define the operator-function
  \begin{equation*}
    (k,\lambda)\mapsto R_\mathcal{M}(k,\la):=\mathcal{J}Z(k,\lambda)
    (I -\mathcal{F}Z(k,\lambda))Q(\la).
  \end{equation*}
  Then, one has
  \begin{enumerate}
  \item $(k,\lambda)\mapsto R_\mathcal{M}(k,\lambda)$ is analytic in
    $\Xi_1$ with values in $B(H^1_a, H^2_{-a}))$;
  \item for $(k,\lambda)\in\Xi_1$, $k\in\R^d$, $\im\lambda^2>0$, there
    exists $\mathcal{H}(k)\subset H^1_a(k)$ such that
    $\overline{{\mathcal H}(k)}=L_2(\Omega)$ and for
    $U\in\mathcal{H}(k)$,
    \begin{equation*}
       R_\mathcal{M} (k, \la) U = (\mathcal{M} (k) - \la)^{-1} U
    \end{equation*}
  \end{enumerate}
\end{Le}
\begin{proof}
  The first property is true because ${\mathcal F}$ is a bounded
  operator from $H^1_{-a}$ to $L_{2, a}$.\\
  To prove the second one, pick $(k,\lambda)\in\Xi_1$ such that
  $k\in\R^d$ and $\im\lambda^2>0$; define
  \begin{equation*}
    {\mathcal H}(k)=({\mathcal M}(k)-\la)H^2_a (k).
  \end{equation*}
  That $\mathcal{H}(k)$ is dense in $L^2(\Omega)$ is a consequence of
  the self-adjointness of $\mathcal{M}$ and the fact that
  $\lambda\not\in\R$.\\
  Let $W\in H^2_a (k)$ and $U=({\mathcal M}-\la){\mathcal J} W$.
  Then, one computes
  \begin{equation}
    \label{eq:38}
    \begin{split}
      R_\mathcal{M}(k,\la)U&=\mathcal{J}Z(k,\lambda)
      (I-\mathcal{F}Z(k,\lambda))Q(\la)({\mathcal M}-\la){\mathcal J}W\\
      &=\mathcal{J}(Z(k,\lambda)-Z(k,\lambda)\mathcal{F}Z(k,\lambda))
      \left(-\Delta_8+\mathcal{V}- \varepsilon\mu\mathcal{J}^{-1}
        (\lambda\mathcal{A}+\lambda^2)\mathcal{J}+\mathcal{F}\right)W\\
      &=\mathcal{J}\left[W+Z(k,\lambda)\mathcal{F}W\right.\\
      &\hskip2cm\left.-Z(k,\lambda)\mathcal{F}Z(k,\lambda)
        \left(-\Delta_8+\mathcal{V}-
          \varepsilon\mu\mathcal{J}^{-1}(\lambda\mathcal{A}+
          \lambda^2)\mathcal{J}+\mathcal{F}\right)W\right]\\
      &=\mathcal{J} (W - Z(k,\lambda)\mathcal{F}
      Z(k,\lambda)\mathcal{F} W)
    \end{split}
  \end{equation}
  where we used (\ref{eq:13}) and (\ref{eq:2.4}).
  Furthermore, one can check that
  \begin{equation*}
   \mathcal{F} Z(k, \la)\mathcal{F} = 0.
  \end{equation*}
  Plugging this into~\eqref{eq:38}, we obtain
  \begin{equation*}
    R_\mathcal{M}(k,\la)U=\mathcal{J}W=(\mathcal{M}(k)-\la)^{-1}U.
  \end{equation*}
  This completes the proof of Lemma~\ref{Lem2.3}.
\end{proof}
\begin{Rem}
  \label{rem:1}
  One presumably has ${\mathcal H} (k) = H^1_a (k)$.
\end{Rem}
\begin{Le}
  \label{Lem2.4}
  Let $G_0$ and $G$ be two Hilbert spaces, $G_0\subset G$, and $G_0^*$
  be a dual space to $G_0$ with respect to the scalar product in $G$.
  Let $B$ be a self-adjoint operator in $G$.  Suppose that $R_B$ is an
  analytic function defined in a complex neighborhood of an interval
  $[\al,\be]$ except at a finite number of points
  $\{\mu_1,\dots,\mu_N\}$, that the values of $R_B$ are in $B (G_0,
  G_0^*)$ and that
  \begin{equation*}
    R_B (\la) \ph = (B - \la)^{-1} \ph \quad \text{if } \im \la > 0, \ph
  \in {\mathcal H}
  \end{equation*}
  where ${\mathcal H}\subset G_0$ is dense in $G$.  Then, the spectrum
  of $B$ in the set $[\al,\be]\setminus\{\mu_1, \dots, \mu_N \}$ is
  absolutely conti\-nuous.  If $\La \subset [\al, \be]$, $\mes\La = 0$
  and $\mu_j \not\in \La$, $j = 1, \dots, N$, then $E_B (\La) = 0$,
  where $E_B$ is the spectral projector of $B$.
\end{Le}
\noindent This lemma is an immediate consequence of Proposition 2 and
equation (18) in section 1.4.5 of~\cite{MR94f:47012}.\\
Now, let $G$ be a Hilbert space, and let $(H(k))_{k\in\C^d}$ be an
analytic family of self-adjoint operators on $G$.  On
$\mathcal{G}=L^2([0,1)^d,G)$, following~\cite{MR58:12429c}, one
defines the self-adjoint operator
\begin{equation*}
  H=\int_{[0,1)^d}^\oplus H (k) dk.
\end{equation*}
The following abstract theorem on the spectrum of the fibered operator
$H$ is based on the Lemma \ref{Lem2.4}.  Its proof repeats the proof
of Theorem 1.1 in~\cite{Fi-Kl:04} although this explicit formulation
is not given there.
\begin{Th}
  \label{thr:4}
  Suppose that there exists a sequence
  %of open sets $U_m\subset \C^{d+1}$ and a sequence
  of analytic functions $f_m:\C^{d+1}\to\C$ such
  that
  \begin{equation*}
    \forall\la\quad\exists k\quad \text{such that}\quad f_m
    (k,\la)\not= 0,
  \end{equation*}
  and the set of real points $(k,\la)$ where $f_m(k,\la)\not=0$ for
  all $m$ can be represented as
  \begin{equation*}
    \R^{d+1}\setminus \bigcup_{m=1}^\infty \{(k,\lambda) : f_m (k,\la)=
    0\} = \bigcup_{j=1}^\infty B_{\er_j} (k_j) \times B_{\er_j} (\la_j).
  \end{equation*}
  Suppose moreover that, for every $j$, there exist
  \begin{itemize}
  \item an analytic scalar function $h_j$ defined in a complex
    neighborhood of $\overline{B_{\er_j} (k_j) \times B_{\er_j} (\la_j)}$
    satisfying property (\ref{eq:2.3});
  \item a Hilbert space $G_j (k)\subset G$, its dual $G^*_j (k)$ with
    respect to the scalar product in $G$, and a set ${\mathcal
      H}_j(k)$ such that
   $${\mathcal H}_j (k)\subset G_j (k)\subset G,\quad
      \overline{{\mathcal H}_j (k)}=G;$$
  \item an analytic $B(G_j,G_j^*)$-valued function $R_j$ defined
    on the set $\{(k,\lambda);\ h_j(k,\la)\not=0\}$ such that for
    $k\in\R^d$, $\im\la>0$, $f\in {\mathcal H}_j(k)$,
    \begin{equation*}
      R_j (k, \la) f = (H(k) - \la)^{-1} f.
    \end{equation*}
  \end{itemize}
  Then, the spectrum of $H$ is purely absolutely continuous.
\end{Th}
\noindent The spectral theory of a class of analytically fibered
operators has been studied in~\cite{MR99b:47030}; their definition of
an analytically fibered operator cannot be used in the present case as
they require the resolvent of the fiber operators to be compact.
\smallpagebreak Theorem~\ref{thr:4} completes the proof of
Theorem~\ref{thr:2} if we take
\begin{equation*}
  G = L_2 (\Omega), \quad H(k) = {\mathcal M} (k),
  \quad H = {\mathcal M}, \quad f_n (k, \la) = (k+n)^2 - \er_0\mu_0 \la^2,
\end{equation*}
use Lemma~\ref{Lem2.3} in a neighborhood of each pair $(k,\la)$ for
which $f_n$ does not vanish, and set
\begin{equation*}
  {\mathcal H}_j(k) = ({\mathcal M} - \la) H^2_{a_j} (k), \quad
  G_j (k) = H^1_{a_j} (k), \quad  H^2_{-a_j} (k)\subset G^*_j (k) =
  H^{-1}_{-a_j} (k), \quad R_j = R_{\mathcal M}.
\end{equation*}

%%% Local Variables:
%%% mode: latex
%%% TeX-master: "maxwell"
%%% End: 